\documentclass[preprint,showpacs,preprintnumbers,nofootinbib]{revtex4}

\usepackage[dvips]{color}% Use color
\usepackage{graphicx}
\usepackage{bbm}

\newcommand{\diag}{{\rm Diag}}
\newcommand{\ev}{{\rm eV}}
\newcommand{\mev}{{\rm MeV}}
\newcommand{\gev}{{\rm GeV}}

\newcommand{\bmx}{\left(\begin{array}}
\newcommand{\emx}{\end{array}\right)}

\begin{document}

\title{Finite quantum corrections to the tribimaximal neutrino mixing}
\author{
Takeshi Araki$^{a)}$\footnote{araki@ihep.ac.cn}, Chao-Qiang
Geng$^{b)}$\footnote{geng@phys.nthu.edu.tw} and Zhi-zhong
Xing$^{a)}$\footnote{xingzz@ihep.ac.cn} } \affiliation{
$^{a)}$Institute of High Energy Physics, Chinese Academy of
Sciences, Beijing 100049, China \\
$^{b)}$Department of Physics,
National Tsing Hua University,
Hsinchu 300, Taiwan}

\begin{abstract}
We calculate {\it finite} quantum corrections to the tribimaximal
neutrino mixing pattern $V^{}_{\rm TB}$ in three generic classes of
neutrino mass models. We show that three flavor mixing angles can
all depart from their tree-level results described by $V^{}_{\rm
TB}$, among which $\theta^{}_{12}$ is most sensitive to such quantum
effects, and the Dirac CP-violating phase can radiatively arise from
two Majorana CP-violating phases. This theoretical scheme offers a
new way to understand why $\theta^{}_{13}$ is naturally small and
how three CP-violating phases are presumably correlated.
\end{abstract}

\pacs{14.60.Pq, 12.60.-i}

\maketitle

Since the discovery of neutrino oscillations \cite{pdg}, many
efforts have been devoted to establishing a new theoretical
framework to accommodate tiny neutrino masses and large flavor
mixing angles. In particular, many studies have tried to parametrize
the Maki-Nakagawa-Sakata-Pontecorvo (MNSP) matrix \cite{PMNS} in
terms of only constant numbers. The most successful parametrization
is known as the tribimaximal (TB) mixing pattern~\cite{TB},
\begin{eqnarray}
V^{}_{\rm TB}= \frac{1}{\sqrt{6}} \bmx{ccc}
 2 & \sqrt{2} & 0 \\
 -1 & \sqrt{2} & -\sqrt{3} \\
 -1 & \sqrt{2} & \sqrt{3}
\emx \Omega \; , \label{eq:TB}
\end{eqnarray}
where $\Omega=\diag\{e^{-i\rho/2},e^{-i\sigma/2},1\}$ includes two
CP-violating phases if three neutrinos are the Majorana particles.
This ansatz predicts $\theta^{\rm TB}_{23}=45^\circ$, $\theta^{\rm
TB}_{12}\simeq 35.26^\circ$ and $\theta^{\rm TB}_{13}=0^\circ$ in
the standard representation of the $3\times 3$ neutrino mixing
matrix \cite{pdg}. So the Dirac CP-violating phase $\delta$ is not
well-defined and there is no CP violation in neutrino oscillations.
Natural realizations of the TB mixing pattern have been explored in
many flavor models~\cite{nADFS}, especially those with an $A_4$
flavor symmetry~\cite{TB-A4}.

However, the present experimental data seem to suggest a small
deviation from the TB mixing. For instance, some hints of
$\theta^{}_{13}>0^\circ$ are claimed in Refs. \cite{Fogli} and
\cite{gfit} at the $1\sigma$ or $2\sigma$ significance level, and
the latest analysis by the KamLAND Collaboration \cite{Kam}
indicates a similar nonzero value of $\theta^{}_{13}$. On the
theoretical side, a deviation from $V^{}_{\rm TB}$ is naturally
expected as there is no good reason for $\theta^{}_{13} =0^\circ$
and CP invariance in neutrino oscillations. Hence it is reasonable
to conjecture that the TB mixing is exact only at the zeroth order
or tree level, and the full flavor mixing matrix $V$ arises from
slight corrections to $V^{}_{\rm TB}$. But at a {\it given} energy
scale such corrections are usually introduced by hand in the absence
of a deeper understanding of why they are small and CP-violating.

In this Letter we look at a new theoretical scheme to account for
possible departures from the TB mixing pattern $V^{}_{\rm TB}$. It
is based on {\it finite} quantum (loop) corrections to $V^{}_{\rm
TB}$. Once $V^{}_{\rm TB}$ is given at the tree level and at an
arbitrary energy scale, we show that a specific and small correction
to $V^{}_{\rm TB}$ can naturally be obtained from finite quantum
effects. Nonzero $\theta^{}_{13}$ and $\delta$ can therefore be
generated, and they may serve for the discriminator of this
theoretical scheme from other neutrino mass models.

We start by considering a generic model which can predict the TB
neutrino mixing $V^{}_{\rm TB}$ before quantum corrections are taken
into account. In this case the tree-level Majorana neutrino mass
matrix can be expressed in terms of its three eigenvalues
($\lambda^{}_1$, $\lambda^{}_2$ and $\lambda^{}_3$):
\begin{eqnarray}
M_\nu^0 =
 \frac{\lambda^{}_1 e^{i\rho}}{6}
 \left(\begin{array}{ccc}
  4 & -2 & -2 \\
  -2 & 1 & 1 \\
  -2 & 1 & 1
 \end{array}\right)
+\frac{\lambda^{}_2 e^{i\sigma}}{3}
 \left(\begin{array}{ccc}
  1 & 1 & 1 \\
  1 & 1 & 1 \\
  1 & 1 & 1
 \end{array}\right)
+\frac{\lambda^{}_3}{2}
 \left(\begin{array}{ccc}
  0 & 0 & 0 \\
  0 & 1 & -1 \\
  0 & -1 & 1
 \end{array}\right) \label{eq:mtb}
\end{eqnarray}
in the basis where the flavor eigenstates of three charged leptons
are identified with their mass eigenstates. Note that $\rho$ and
$\sigma$ are the Majorana CP-violating phases defined in
Eq.~(\ref{eq:TB}). Let us explicitly examine how {\it finite}
quantum corrections modify $V^{}_{\rm TB}$. To do so, we take the
form of the full neutrino mass matrix as
\begin{eqnarray}
M^{}_\nu = M_\nu^0 + \Delta M^{}_\nu \, ,
\end{eqnarray}
where $\Delta M^{}_\nu$ arises from some loop corrections. Here we
have assumed that $\Delta M^{}_\nu$ is determined only by $M_\nu^0$
and three charged-lepton masses ($m^{}_e$, $m^{}_\mu$ and
$m^{}_\tau$), in order to make the theory as predictable as
possible. We do not introduce any new Yukawa couplings. Note that
the type-II seesaw mechanism~\cite{htm} may induce not only the
tree-level neutrino mass term but also the desired quantum
corrections due to the singly- and doubly-charged components of the
triplet scalar and moreover, it does not suffer from any intrinsic
{\it non-unitary} effects on the MNSP matrix~\cite{non-uni}.
Regardless of any model details, we proceed to analyze three generic
classes of loop corrections:
\begin{figure}[t]
\begin{center}
\includegraphics*[width=0.4\textwidth]{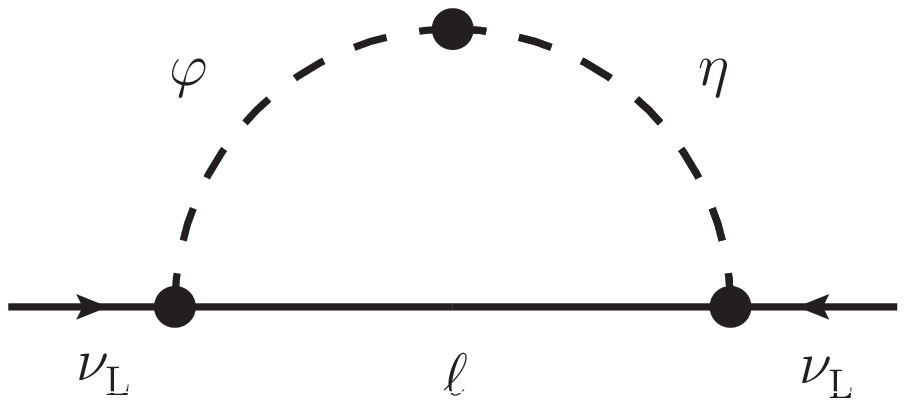}
\vspace{-0.3cm} \caption{\footnotesize {A one-loop neutrino mass
operator, where $\varphi$ and $\eta$ (or $\ell$) stand for new
scalars (or the charged leptons in the standard model). }
}\vspace{1.0cm}\label{fig:1loop}
\includegraphics*[width=0.4\textwidth]{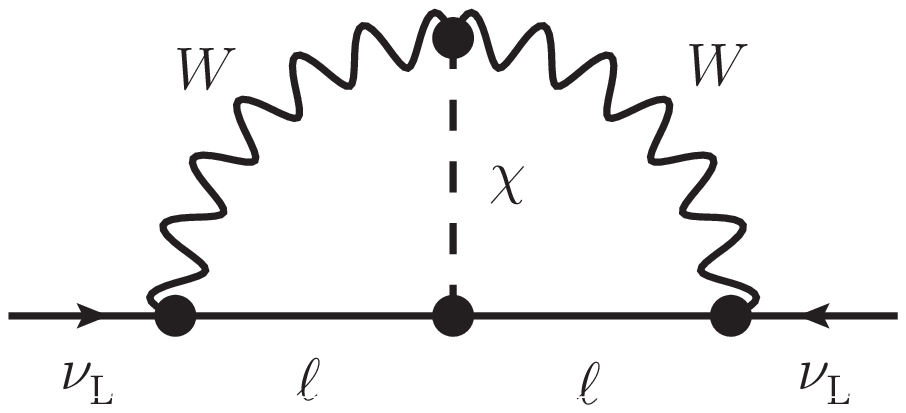}
\includegraphics*[width=0.4\textwidth]{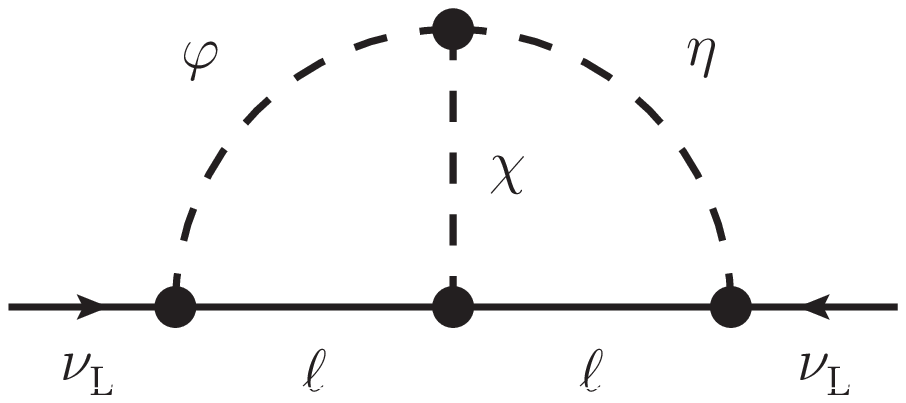}
\vspace{-0.3cm}
\caption{\footnotesize Two-loop neutrino mass operators mediated by
two $W$ bosons (left panel) and three scalars $\varphi$, $\eta$ and $\chi$
(right panel). }\label{fig:2loop}
\end{center}
\end{figure}
\begin{itemize}
\item Class-I
\begin{eqnarray}
(\Delta M_\nu^{})_{\alpha\beta}^{} =
\frac{(M_\nu^0)_{\alpha\beta}^{} m_\beta^2 + m_\alpha^2
(M_\nu^0)_{\alpha\beta}^{}}{v^2} \times I^{\rm loop} \; ,
\label{eq:mn-I}
\end{eqnarray}
where the Greek subscripts $\alpha$ and $\beta$ run over $e$, $\mu$
and $\tau$, $v \simeq 246\ \gev$ is the vacuum expectation value of
the standard-model Higgs field, and $I^{\rm loop}$ denotes a
dimensionless function from the loop integral. An example of this
class of quantum corrections can be displayed by the one-loop
Feynman diagram in FIG.~\ref{fig:1loop}~\cite{zee}.

\item Class-II
\begin{eqnarray}
(\Delta M_\nu^{})_{\alpha\beta}^{} = \frac{m_\alpha^{}
(M_\nu^0)_{\alpha\beta}^{} m_\beta^{}}{v^2} \times I^{\rm loop} \; .
\label{eq:mn-II}
\end{eqnarray}
Such a correction term may emerge from the two-loop Feynman diagram
shown in FIG.~\ref{fig:2loop} (left panel), where $\chi$ corresponds
to a doubly-charged scalar~\cite{geng}.

\item Class-III
\begin{eqnarray}
(\Delta M_\nu^{})_{\alpha\beta}^{} =
\frac{(\tilde{M}_\nu^0)_{\alpha\xi}^{} m_\xi^{}
(M_\nu^0)^{*}_{\xi\zeta}
       m_\zeta^{} (\tilde{M}_\nu^0)_{\zeta \beta}^{}}{v^2}
\times I^{\rm loop} \; ,
\label{eq:mn-III}
\end{eqnarray}
where $\tilde{M}_\nu^0 \equiv M_\nu^0/(1~\ev)$, and the Greek
subscripts $\alpha$, $\beta$, $\xi$ and $\zeta$ run over $e$, $\mu$
and $\tau$. This correction term may be generated through a two-loop
diagram mediated by three scalars as shown in FIG.~\ref{fig:2loop}
(right panel), where $\varphi$ and $\eta$ are singly-charged scalars
and $\chi$ denotes a doubly-charged scalar in the Zee-Babu
model~\cite{zee-babu}.
\end{itemize}
In each case one can calculate three neutrino masses $m^{}_i$ (for
$i=1,2,3$) and the flavor mixing matrix $V$ by diagonalizing the full
neutrino mass matrix $M^{}_\nu =M^0_\nu + \Delta M^{}_\nu$.

Since quantum corrections are expected to be very small, they may
serve for small perturbations to the tree-level terms. After doing
some perturbation calculations with the help of Eqs. (2)---(6), we
arrive at the neutrino masses
\begin{eqnarray}
m_1^{}\simeq|\lambda_1^{} e_{}^{i\rho} + P_{11}^{}| \; , ~~
m_2^{}\simeq|\lambda_2^{} e_{}^{i\sigma} + P_{22}^{}| \; , ~~
m_3^{}\simeq|\lambda_3^{} + P_{33}^{}| \; ,
\label{eq:m}
\end{eqnarray}
where $P^{}_{ij} \equiv (V_{\rm TB}^T \Delta M_\nu^{} V^{}_{\rm
TB})^{}_{ij}$ for each class of loop corrections. In deriving the
expression of $V$ we find that $m^{}_i \simeq \lambda^{}_i$ is
actually a good approximation even in the special case of
$\lambda^{}_1=0$ or $\lambda^{}_3=0$. So we obtain the following
modified neutrino mixing angles in the standard parametrization of
$V$:
\begin{eqnarray}
&&\sin\theta^{}_{13}
\simeq
\frac{1}{\sqrt{3}}
\left|
 \frac{\sqrt{2} P_{13}^{}}{m^{}_3-m^{}_1 e^{i\rho}}
+\frac{P_{23}^{}}{m^{}_3-m^{}_2 e^{i\sigma}}
\right| \; , \nonumber \\
%\label{eq:s13}\\
%
&&\tan\theta^{}_{23}
\simeq
\left|
1 + \frac{2}{\sqrt{3}}
  \left[
   \frac{P_{13}^{}}{m^{}_3-m^{}_1 e^{i\rho}}
 -\frac{\sqrt{2}P_{23}^{}}{m^{}_3-m^{}_2 e^{i\sigma}}
  \right]\right| \; , ~~ \nonumber \\
%\label{eq:t23}\\
%
&&\tan\theta^{}_{12}
\simeq
\frac{1}{\sqrt{2}}
\left|
1 + \frac{3}{\sqrt{2}}
\frac{P_{12}^{}}{m^{}_2 e^{i\sigma}-m^{}_1 e^{i\rho}}
\right| \; . \label{eq:t12}
\end{eqnarray}
More explicit expressions of these mixing parameters are shown in
Appendix A for each class of loop corrections. We see that the
quantum effects on three mixing angles are proportional to the loop
function $I^{\rm loop}$ and depend crucially upon two Majorana
CP-violating phases and the near degeneracy of three neutrino
masses. In view of $\Delta m^2_{21} \simeq 7.6 \times 10^{-5} ~{\rm
eV}^2$ and $\Delta m^2_{31} \simeq \Delta m^2_{32} \simeq \pm 2.4
\times 10^{-3} ~{\rm eV}^2$ \cite{gfit}, we conclude that
$\theta^{}_{12}$ is in general more sensitive to radiative
corrections than $\theta^{}_{13}$ and $\theta^{}_{23}$. The maximal
departure of $\theta^{}_{12}$ from $\theta^{\rm TB}_{12}$ takes
place when $\sigma = \rho =0^\circ$ (or $\pm 180^\circ$). More
interestingly, the smallest neutrino mixing angle $\theta^{}_{13}$
becomes nonzero thanks to finite loop corrections. Similar
observations have been made when one investigates the running
behaviors of three neutrino mixing angles from a superhigh seesaw
scale to the electroweak scale (or vice versa) by means of the
one-loop renormalization-group equations (RGEs) \cite{RGE}. The main
difference between the quantum effect revealed in Eq. (8) and the
RGE running effect is that the latter primarily describes the {\it
evolution} of relevant physical quantities with {\it different}
energy scales. Note that a nontrivial value of the Dirac
CP-violating phase $\delta$ can be generated together with
$\theta^{}_{13}$, leading to leptonic CP violation in neutrino
oscillations whose strength is measured by the rephasing-invariant
Jarlskog parameter \cite{jarl}
\begin{eqnarray}
J =s_{12}^{}c_{12}^{}s_{23}^{}c_{23}^{}s_{13}^{}c_{13}^2\sin\delta
\simeq \frac{1}{3\sqrt{6}} \left[ {\rm Im}\left(\frac{\sqrt{2}
P_{13}^{}}{m^{}_3-m^{}_1e^{i\rho}}\right) + {\rm Im}\left(\frac{
P_{23}^{}}{m^{}_3-m^{}_2e^{i\sigma}}\right) \right] \; ,
\end{eqnarray}
where $s^{}_{ij} \equiv \sin\theta^{}_{ij}$ and $c^{}_{ij} \equiv
\cos\theta^{}_{ij}$. This result makes it transparent that $\delta$
arises from two Majorana CP-violating phases $\rho$ and $\sigma$ via
quantum corrections, analogous to the radiative generation of
$\delta$ from $\rho$ and $\sigma$ via the RGE running effects
\cite{RGE,Luo}.

Let us add some comments on the finite quantum effects obtained in
Eqs. (8) and (9) as compared with the corresponding RGE running
effects on three neutrino mixing angles and the Jarlskog parameter
\cite{RGE,Luo}. Both of them are suppressed by the factors
$m^2_\alpha/v^2$ (for $\alpha =e, \mu, \tau$). The smallness of
finite quantum corrections is also attributed to the smallness of
the loop function $I^{\rm loop}$, whereas the RGE running effects
might more or less be enhanced by a reasonably large logarithm
$\ln(\mu/\mu^{}_0)$ if the initial ($\mu^{}_0$) and final ($\mu$)
energy scales are considerably different from each other. In Ref.
\cite{Bando} it has been noticed that the RGE running effects are
usually more significant than the finite quantum corrections, if
they originate from the same Feynman diagrams and if $\mu/\mu^{}_0
\gg 1$ (or $\mu/\mu^{}_0 \ll 1$) holds. Furthermore, relatively
strong RGE running effects may appear if the so-called seesaw
threshold effects are taken into account in some neutrino mass
models \cite{RGE2}. Since the size of $I^{\rm loop}$ and seesaw
threshold effects are strongly model-dependent, a quantitative
comparison between two kinds of quantum effects under discussion can
only be made in a specific model and is apparently beyond the scope
of this Letter. We shall present such a comparison elsewhere
\cite{Araki}. But let us stress that in both cases the roles of two
Majorana CP-violating phases are quite similar, and so is the
dependence of quantum corrections on the absolute neutrino mass
scale. In particular, the fact that $\delta$ can be radiatively
generated from $\rho$ and $\sigma$ implies that they are all of the
Majorana nature although $\delta$ is usually called the Dirac
CP-violating phase.
\begin{figure}[t]
\begin{center}
\includegraphics*[width=0.48\textwidth]{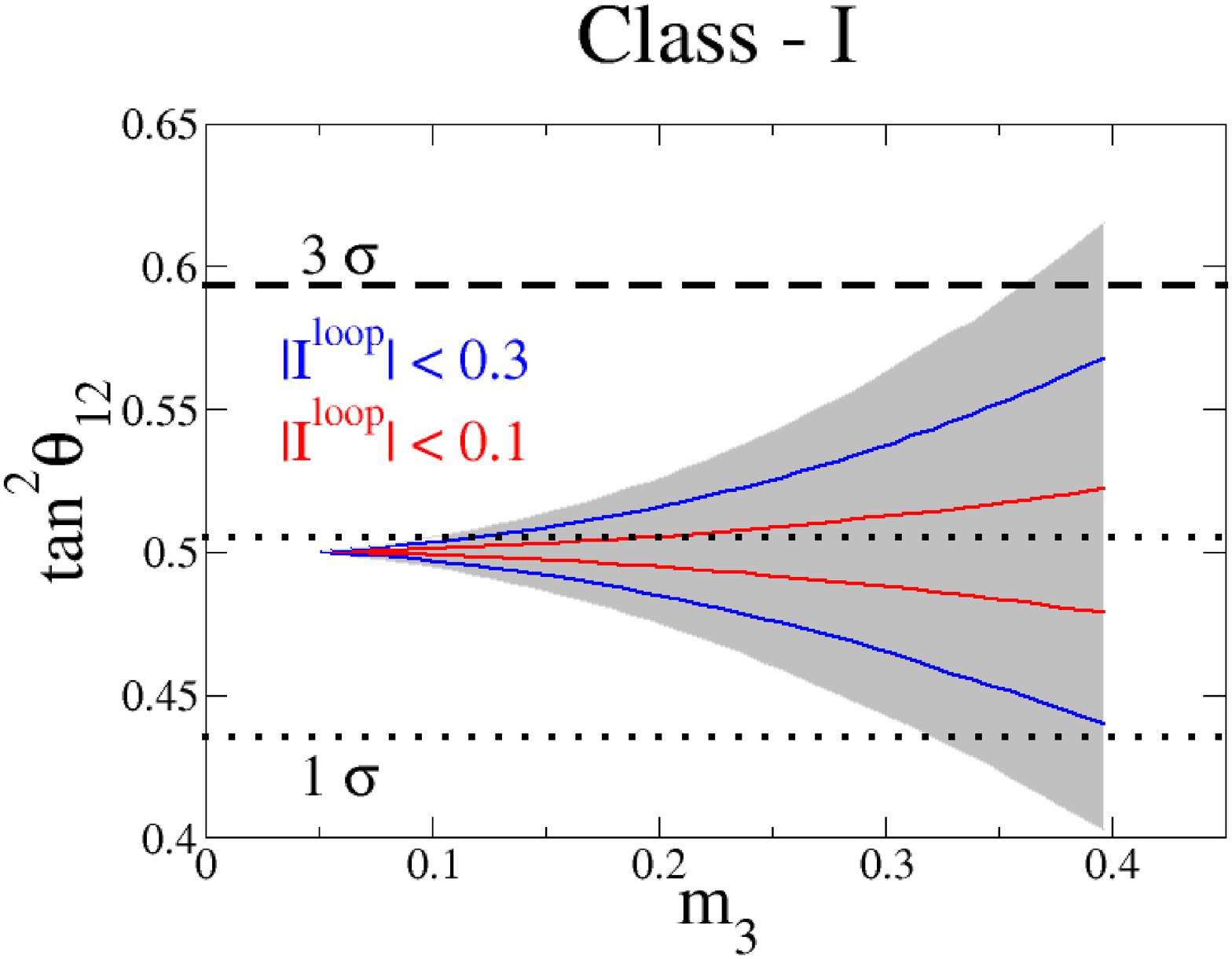}
\includegraphics*[width=0.48\textwidth]{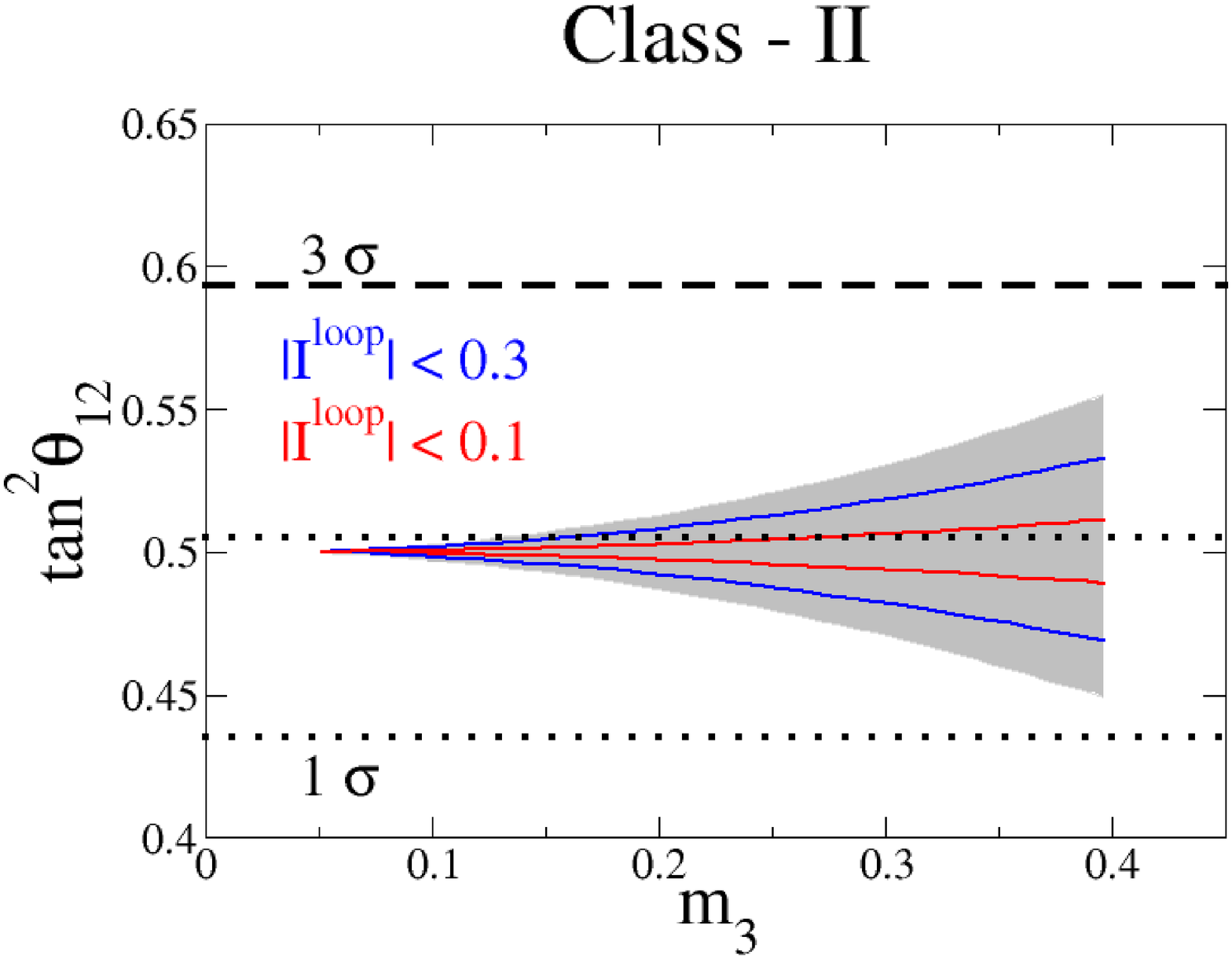}
\vspace{0.8cm}\\
\includegraphics*[width=0.48\textwidth]{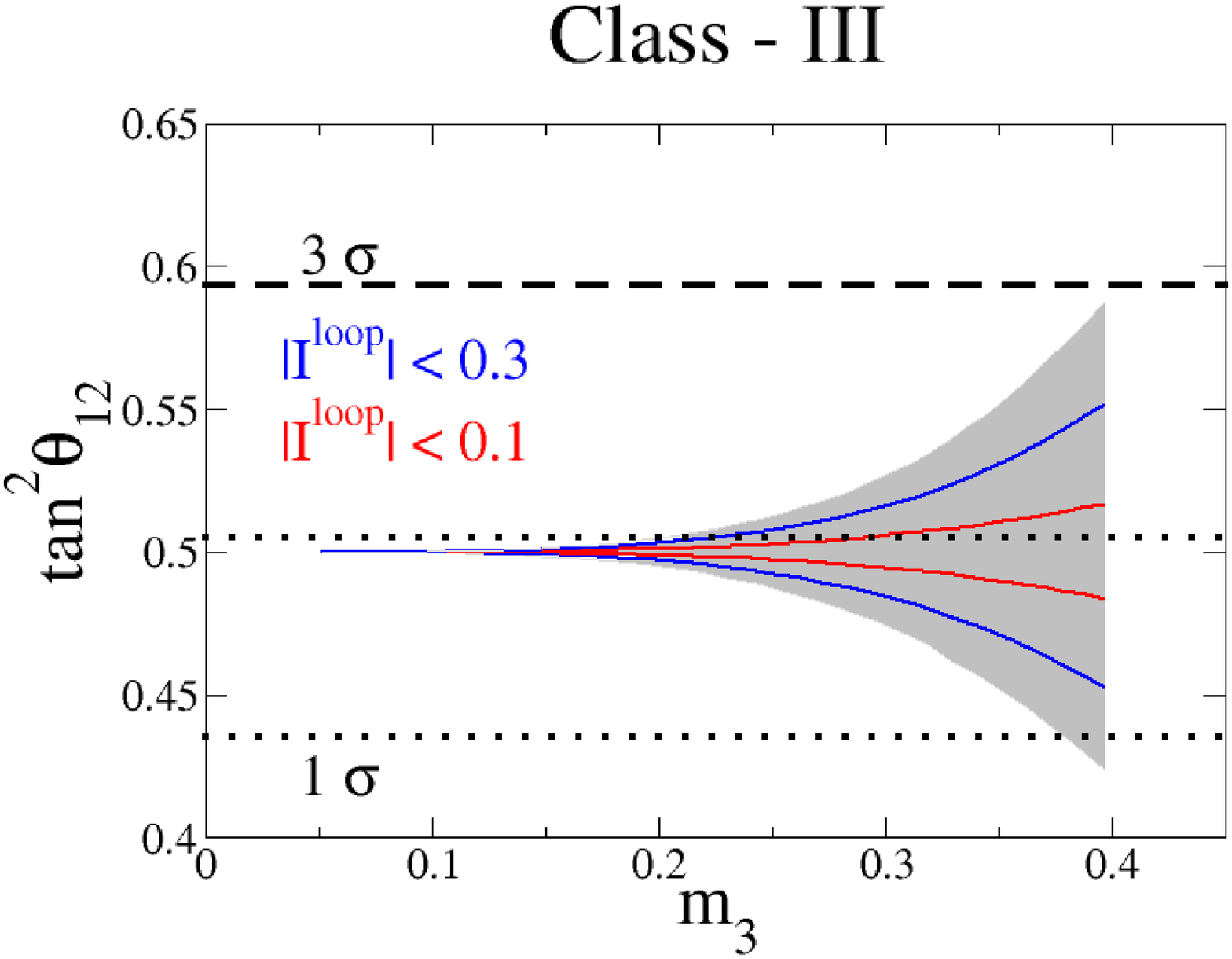}
\caption{\footnotesize $\tan^2\theta^{}_{12}$ as a function of
$m^{}_3$ in three classes of quantum corrections with $\Delta
m^2_{31} >0$. Here the shaded area means the whole allowed region
with respect to $|I^{\rm loop}|<0.5$; the blue (or red) lines show
the boundaries of the allowed region with $|I^{\rm loop}|<0.3\ \ev$
(or $0.1\ \ev$); and the dotted (or dashed) lines signify the
$1\sigma$ (or $3\sigma$) bounds \cite{gfit}. } \label{fig:t12-nh}
\end{center}
\end{figure}

We numerically illustrate the quantum effects on three neutrino
mixing angles at the electroweak scale, where the values of three
charged-lepton masses read $m^{}_e = 0.486\ \mev$, $m^{}_\mu =
102.718\ \mev$ and $m^{}_\tau = 1746.24\ \mev$ \cite{xzz-mass}. The
best-fit values of two neutrino mass-squared differences with the
$1\sigma$ errors are $\Delta m_{21}^2 = (7.59 \pm 0.20) \times
10^{-5}\ \ev^2$ and $\Delta m_{31}^2 = (-2.36 \pm 0.11) \times
10^{-3}\ \ev^2$ or $(+2.46 \pm 0.12) \times 10^{-3}\ \ev^2$
\cite{gfit}. In addition, a relatively generous upper limit on the
sum of three neutrino masses is $m^{}_1 + m^{}_2 + m^{}_3 < 1.19\
\ev$ extracted from current cosmological observational data
\cite{Mn-sum}. Assuming $|I^{\rm loop}| \lesssim 0.5$ and allowing
$\rho$ and $\sigma$ to vary between $0^\circ$ and $360^\circ$, we
calculate $\tan^2\theta_{12}$ and then plot our numerical result in
FIG. \ref{fig:t12-nh}, where only the $\Delta m^2_{31} >0$ case is
taken into account because the result for the $\Delta m^2_{31} <0$
case is not very different. We observe that larger values of
$m^{}_3$ give rise to larger magnitudes of $P^{}_{ij} \equiv (V_{\rm
TB}^T \Delta M_\nu^{} V^{}_{\rm TB})^{}_{ij}$ (for $i,j =1,2,3$) and
thus larger radiative corrections to three neutrino mixing angles.
But only $|\tan^2\theta_{12}-1/2|$ is numerically appreciable. We
find $|\tan^2\theta_{23}-1|<{\cal O}(10^{-3})$,
$\sin^2\theta_{13}<{\cal O}(10^{-6})$ and $J<{\cal O}(10^{-4})$ for
each class of loop corrections, implying that $\theta^{}_{23}$,
$\theta^{}_{13}$ and $\delta$ are not very sensitive to the
loop-induced quantum effects. If a precision measurement of neutrino
oscillations establishes a significant deviation of $\theta^{}_{23}$
from $45^\circ$ and (or) $\theta^{}_{13}$ from $0^\circ$, then the
departure of $V$ from $V^{}_{\rm TB}$ must mainly originate from a
different mechanism \cite{Rodejohann}. But this statement is only
valid for the $|I^{\rm loop}| \lesssim 1$ case under discussion,
which seems to be a natural expectation in model building. If
$|I^{\rm loop}| \sim {\cal O}(10)$ were allowed, relatively larger
quantum corrections to three mixing angles would be expected. We
shall examine whether this case is possible or not in a specific
flavor model elsewhere.

In summary, we have calculated {\it finite} quantum corrections to
the TB neutrino mixing pattern and discussed the generation of
nonzero $\theta^{}_{13}$ and $\delta$ in this way for three generic
classes of neutrino mass models. Among three mixing angles,
$\theta^{}_{12}$ is found to be most sensitive to such quantum
effects. Similar behaviors have been observed in the study of RGE
running effects on neutrino mixing parameters. This theoretical
approach provides a new possibility of understanding why
$\theta^{}_{13}$ is naturally small and how the Dirac and Majorana
CP-violating phases are presumably correlated, in particular when
flavor symmetries are taken as a good starting point of view for
model building so as to derive the most favored neutrino mixing
scheme such as the TB mixing. We stress that all the neutrino mass
models at a given energy scale should carefully take into account
the quantum effects on their tree-level results. Such effects can be
very important in some cases as we have demonstrated, and they are
even accessible in a variety of precision neutrino experiments in
the near future.

\begin{acknowledgments}
The work of T.A. and Z.Z.X. was supported in part by the National
Natural Science Foundation of China under Grant No. 10875131. C.Q.G.
was partially supported  by the National Science Council under Grant
No. NSC-98-2112-M-007-008-MY3 and the National Tsing Hua University
under the Boost Program No. 97N2309F1.
\end{acknowledgments}

\vspace{1cm}

\appendix\section{}

Here let us write out the explicit expressions of Eqs. (8) and (9)
for each class of loop corrections. Neglecting those small terms
proportional to $m^2_e/v^2$ or $m^2_\mu/v^2$ as a good
approximation, we obtain
\begin{eqnarray}
&&\sin\theta^{}_{13} \simeq \frac{1}{3\sqrt{2}}\frac{m_\tau^2}{v^2}
\left|
 \frac{m^{}_3+m^{}_1 e^{i\rho}}{m^{}_3-m^{}_1 e^{i\rho}}
-\frac{m^{}_3+m^{}_2 e^{i\sigma}}{m^{}_3-m^{}_2 e^{i\sigma}}
\right| I^{\rm loop} \nonumber \\
&&\hspace{1.3cm} \simeq \frac{\sqrt{2}}{3}\frac{m_\tau^2}{v^2}
\left[\frac{m_3^2\left[m_1^2+m_2^2-2m^{}_1
m^{}_2\cos(\rho-\sigma)\right]}
     {\left[m_1^2+m_3^2-2m^{}_1 m^{}_3\cos\rho\right]
      \left[m_2^2+m_3^2-2m^{}_2 m^{}_3\cos\sigma\right]
      }\right]^{1/2}
     I^{\rm loop} \; , ~~~~ \nonumber\\
&&\tan\theta^{}_{23} \simeq \left| 1 - \frac{m_\tau^2}{3v^2}
  \left[
   \frac{m^{}_3+m^{}_1 e^{i\rho}}{m^{}_3-m^{}_1 e^{i\rho}}
 +2\frac{m^{}_3+m^{}_2 e^{i\sigma}}{m^{}_3-m^{}_2 e^{i\sigma}}
  \right]I^{\rm loop}
\right| \nonumber\\
&&\hspace{1.3cm}
\simeq
1 -
\frac{m_\tau^2}{3v^2}
\left[
\frac{m_3^2-m_1^2}
     {m_1^2+m_3^2-2m^{}_1 m^{}_3\cos\rho}
+ \frac{2(m_3^2-m_2^2)}
     {m_2^2+m_3^2-2m^{}_2 m^{}_3\cos\sigma}
\right]I^{\rm loop} \; , \nonumber \\
&&\tan\theta^{}_{12} \simeq \frac{1}{\sqrt{2}} \left| 1 -
\frac{m_\tau^2}{2v^2} \frac{m^{}_2e^{i\sigma}+m^{}_1e^{i\rho}}
     {m^{}_2e^{i\sigma}-m^{}_1e^{i\rho}}I^{\rm loop}
\right| \nonumber\\
&&\hspace{1.3cm}
\simeq
\frac{1}{\sqrt{2}}
\left[
1 - \frac{m_\tau^2}{2v^2}
\frac{m_2^2-m_1^2}
     {m_2^2+m_1^2 - 2m^{}_1 m^{}_2\cos(\rho-\sigma)}I^{\rm loop}
\right] \; , \nonumber \\
&&J\simeq
\frac{m_\tau^2}{9v^2}
\left[
 \frac{m^{}_2 m^{}_3\sin\sigma}
      {m_2^2+m_3^2-2m^{}_2m^{}_3\cos\sigma}
-\frac{m^{}_1 m^{}_3\sin\rho}
      {m_1^2+m_3^2-2m^{}_1m^{}_3\cos\rho}
\right]I^{\rm loop}
\end{eqnarray}
for Class-I;
\begin{eqnarray}
&&\sin\theta^{}_{13} \simeq \frac{1}{18\sqrt{2}}\frac{m_\tau^2}{v^2}
\left|
 \frac{M}{m^{}_3-m^{}_1e^{i\rho}}
-\frac{M}{m^{}_3-m^{}_2e^{i\sigma}}
\right| I^{\rm loop} \nonumber\\
&&\hspace{1.3cm} \simeq \frac{\sqrt{2}}{36}\frac{m_\tau^2}{v^2}
\left[\frac{|M|^2 \left[m_1^2+m_2^2-2m^{}_1
m^{}_2\cos(\rho-\sigma)\right]}
     {\left[m_1^2+m_3^2-2m^{}_1 m^{}_3\cos\rho\right]
      \left[m_2^2+m_3^2-2m^{}_2 m^{}_3\cos\sigma\right] }
     \right]^{1/2} I^{\rm loop} \; , \nonumber \\
&&\tan\theta^{}_{23} \simeq \left| 1 - \frac{m_\tau^2}{18v^2}
  \left[
   \frac{M}{m^{}_3-m^{}_1e^{i\rho}}
 +2\frac{M}{m^{}_3-m^{}_2e^{i\sigma}}
  \right]I^{\rm loop}
\right|  \nonumber\\
&&\hspace{1.3cm} \simeq 1 - \frac{m_\tau^2}{18v^2} \left[
\frac{3m_3^2-m_1^2-2m^{}_1m^{}_3\cos\rho
       +2m^{}_2m^{}_3\cos\sigma-2m^{}_1m^{}_2\cos(\rho-\sigma)}
     {m_1^2+m_3^2-2m^{}_1 m^{}_3\cos\rho}
\right. \nonumber\\
&& \hspace{3.0cm} \left.+2
\frac{3m_3^2-2m_2^2-m^{}_2m^{}_3\cos\sigma
         +m^{}_1m^{}_3\cos\rho-m^{}_1m^{}_2\cos(\rho-\sigma)}
     {m_2^2+m_3^2-2m^{}_2 m^{}_3\cos\sigma}
\right]I^{\rm loop} \; , \nonumber \\
&&\tan\theta^{}_{12} \simeq \frac{1}{\sqrt{2}} \left| 1 -
\frac{m_\tau^2}{12v^2} \frac{M}
     {m^{}_2e^{i\sigma}-m^{}_1e^{i\rho}}I^{\rm loop}
\right|  \nonumber\\
&&\hspace{1.3cm} \simeq \frac{1}{\sqrt{2}} \left[ 1 -
\frac{m_\tau^2}{12v^2}
\frac{2m_2^2-m_1^2-m^{}_1m^{}_2\cos(\rho-\sigma)
      +3m^{}_2m^{}_3\cos\sigma-3m^{}_1m^{}_3\cos\rho}
     {m_2^2+m_1^2 - 2m^{}_1 m^{}_2\cos(\rho-\sigma)}I^{\rm loop}
\right] \; , \nonumber \\
&&J\simeq
\frac{m_\tau^2}{108v^2}
\left[
 \frac{m^{}_1m^{}_3\sin\rho+5m^{}_2 m^{}_3\sin\sigma
       -m^{}_1m^{}_2\sin(\rho-\sigma)}
      {m_2^2+m_3^2-2m^{}_2m^{}_3\cos\sigma}
\right.\nonumber \\
&&\hspace{2.5cm} \left. -\frac{4m^{}_1
m^{}_3\sin\rho+2m^{}_2m^{}_3\sin\sigma
       +2m^{}_1m^{}_2\sin(\rho-\sigma)}
      {m_1^2+m_3^2-2m^{}_1m^{}_3\cos\rho}
\right]I^{\rm loop}
\end{eqnarray}
for Class-II; and
\begin{eqnarray}
&&\sin\theta^{}_{13} \simeq \frac{1}{18\sqrt{2}}
\frac{m_\tau^2}{v^2} \left|
 \frac{M^* \tilde{m}^{}_3\tilde{m}^{}_1e^{i\rho}}
      {m^{}_3-m^{}_1e^{i\rho}}
-\frac{M^* \tilde{m}^{}_3\tilde{m}^{}_2e^{i\sigma}}
      {m^{}_3-m^{}_2e^{i\sigma}}
\right| I^{\rm loop}\nonumber \\
&&\hspace{1.3cm} \simeq \frac{\sqrt{2}}{36} \frac{m_\tau^2
m^{}_3\tilde{m}^{}_3}{v^2} \left[ \frac{|M|^2
\left[\tilde{m}_1^2+\tilde{m}_2^2
      -2\tilde{m}^{}_1 \tilde{m}^{}_2\cos(\rho-\sigma)\right]}
     {\left[m_1^2+m_3^2-2m^{}_1 m^{}_3\cos\rho\right]
      \left[m_2^2+m_3^2-2m^{}_2 m^{}_3\cos\sigma\right] }
     \right]^{1/2} I^{\rm loop} \; , \nonumber \\
&&\tan\theta^{}_{23} \simeq \left| 1 - \frac{m_\tau^2}{18v^2}
  \left[
   \frac{M^* \tilde{m}^{}_3\tilde{m}^{}_1e^{i\rho}}
        {m^{}_3-m^{}_1e^{i\rho}}
 +2\frac{M^* \tilde{m}^{}_3\tilde{m}^{}_2e^{i\sigma}}
        {m^{}_3-m^{}_2e^{i\sigma}}
  \right]I^{\rm loop}
\right| \nonumber \\
&&\hspace{1.3cm} \simeq 1 - \frac{m_\tau^2 \tilde{m}^{}_3}{18v^2}
\left[\tilde{m}^{}_1
\frac{3m_3^2\cos\rho-m_1^2\cos\rho-2m^{}_1m^{}_3
       +2m^{}_2m^{}_3\cos(\rho-\sigma)-2m^{}_1m^{}_2\cos\sigma}
     {m_1^2+m_3^2-2m^{}_1 m^{}_3\cos\rho}
\right. \nonumber \\
&& \hspace{3.0cm} \left.+2 \tilde{m}^{}_2
\frac{3m_3^2\cos\sigma-2m_2^2\cos\sigma-m^{}_2m^{}_3
         +m^{}_1m^{}_3\cos(\rho-\sigma)-m^{}_1m^{}_2\cos\rho}
     {m_2^2+m_3^2-2m^{}_2 m^{}_3\cos\sigma}
\right]I^{\rm loop} \; , \nonumber \\
&&\tan\theta^{}_{12} \simeq \frac{1}{\sqrt{2}} \left| 1 -
\frac{m_\tau^2}{12v^2} \frac{M^* \tilde{m}^{}_1\tilde{m}^{}_2
e^{i(\rho+\sigma)}}
     {m^{}_2e^{i\sigma}-m^{}_1e^{i\rho}}I^{\rm loop}
\right|  \nonumber \\
&&\hspace{1.3cm} \simeq \frac{1}{\sqrt{2}} \left[ 1 - \frac{m_\tau^2
\tilde{m}^{}_1\tilde{m}^{}_2}{12v^2}
\right. \nonumber \\
&&\left. \hspace{3cm}\times
\frac{2m_2^2\cos(\rho-\sigma)-m_1^2\cos(\rho-\sigma)-m^{}_1m^{}_2
      +3m^{}_2m^{}_3\cos\rho-3m^{}_1m^{}_3\cos\sigma}
     {m_2^2+m_1^2 - 2m^{}_1 m^{}_2\cos(\rho-\sigma)} I^{\rm loop}
\right] \; , \nonumber \\
&&J\simeq- \frac{m_\tau^2 \tilde{m}^{}_3}{108v^2}
\left[\tilde{m}^{}_2
 \frac{m^{}_1m^{}_3\sin(\rho-\sigma)-2m_2^2\sin\sigma
       -3m_3^2\sin\sigma-m^{}_1m^{}_2\sin\rho}
      {m_2^2+m_3^2-2m^{}_2m^{}_3\cos\sigma}
\right.\nonumber \\
&&\hspace{3.0cm} \left.
+\tilde{m}^{}_1\frac{m_1^2\sin\rho+3m_3^2\sin\rho
       +2m^{}_2m^{}_3\sin(\rho-\sigma)+2m^{}_1m^{}_2\sin\sigma}
      {m_1^2+m_3^2-2m^{}_1m^{}_3\cos\rho}
\right]I^{\rm loop}
\end{eqnarray}
for Class-III, where $M\equiv
m^{}_1e^{i\rho}+2m^{}_2e^{i\sigma}+3m^{}_3$ and $\tilde{m}^{}_i
\equiv m^{}_i/(1~\ev)$ have been defined to simplify the expressions
in Eq. (A3) to some extent.

\end{document}